\documentclass[ste,twoside]{stefano}

\usepackage{amsmath}
\usepackage{amssymb}
\usepackage{graphics}
\usepackage{rotating}
\usepackage{cite}
\usepackage{color}

\def\makeheadbox{{%
\hbox to0pt{\vbox{\baselineskip=10dd\hrule\hbox
to\hsize{\vrule\kern3pt\vbox{\kern3pt \hbox{  {\sc The European
Physical Journal C} {\bf 51}, 241-247 (2007)} \hbox{ {\sc
{\color{blue}{dma}}[{\color{black}{imecc}}]{\color{red}{UniCamp}}
} \hspace*{10.4cm} {\color{blue}{$\boldsymbol{\Sigma \delta
\Lambda}$}} }
\kern3pt}\hfil\kern3pt\vrule}\hrule}%
\hss}}}

\def\0{\mbox{\tiny $0$}}
\def\1{\mbox{\tiny $1$}}
\def\2{\mbox{\tiny $2$}}
\def\3{\mbox{\tiny $3$}}
\def\4{\mbox{\tiny $4$}}
\def\5{\mbox{\tiny $5$}}
\def\6{\mbox{\tiny $6$}}
\def\7{\mbox{\tiny $7$}}
\def\8{\mbox{\tiny $8$}}
\def\9{\mbox{\tiny $9$}}
\def\I{\mbox{\tiny $I$}}
\def\II{\mbox{\tiny $II$}}
\def\III{\mbox{\tiny $III$}}

\def\NR{\mbox{\tiny NR}}

\def\pl{\mbox{\tiny $+$}}

\def\inc{\mbox{\tiny inc}}

\def\tra{\mbox{\tiny tra}}
\begin{document}
%

\title{\large  DIRAC EQUATION STUDIES IN THE TUNNELLING ENERGY ZONE}

\author{
Stefano De Leo\inst{1}
\and Pietro P. Rotelli\inst{2} }

\institute{
Department of Applied Mathematics, State University of Campinas\\
PO Box 6065, SP 13083-970, Campinas, Brazil\\
{\em deleo@ime.unicamp.br}
\and
Department of Physics, INFN, University of Lecce\\
PO Box 193, 73100, Lecce, Italy\\
{\em rotelli@le.infn.it}
}


\date{Submitted:  {\em September, 2006}. Revised version: {\em February,
2007}.}


\abstract{We investigate the tunnelling zone $V_{\0}-m<E<V_{\0}+m$ for a
one-dimensional potential within the Dirac equation. We find the appearance
of superluminal transit times akin to the Hartman effect.}



\PACS{ {03.65.Pm\, ,}{} {03.65.Xp\, .}{}}







\titlerunning{\sc dirac equation and tunnelling phenomena}

\maketitle


\section*{I. INTRODUCTION}

In several recent articles, we have investigated in some detail
one-dimensional electrostatic potentials by means of both the Schr\"odinger
\cite{BouS,DifS,TunS} and the Dirac equation \cite{BouD,DifD,KleD}. Several
original phenomena have been observed, such as the transition from
resonance phenomena to multiple (infinite) peak formation \cite{DifD}
equivalent to a shift from wave-like to particle behavior in the barrier
diffusion zone $E>V_{\0}+m$, where $V_{\0}$ is the barrier height, $E$ one
of the wave packet energies and $m$ the particle mass. We have also
investigated the compatibility of the barrier results with the Klein
paradox \cite{K1,K2,K3}, when $m<E<V_{\0}-m$. In this latter case, we have
noted the existence of dynamic localized states with a continuous spectrum
\cite{KleD}. These states are the nearest approximation to the bound states
of Schr\"odinger or of Dirac in the evanescent energy zone considered in
this paper.

The evanescent zone is the last energy zone we need to consider to complete
our analysis. It is given by $V_{\0}-m<E<V_{\0}+m$ ($V_{\0}>2m$) or
$m<E<V_{\0}+m$ ($V_{\0}<2m$). It has evanescent (real exponential) space
forms in the barrier region. For a well potential, it is just such forms
that give rise to discrete bound states. In this paper, we shall
concentrate upon the single barrier potential and hence complete our
analysis for this elementary structure. The evanescent stationary solutions
become dynamic if instead of plane-waves we work with incoming wave
packets. Then, the particles within the classically forbidden region are
measurable only for a finite time, the time of transition from an incoming
wave packet to reflected/transmitted wave packets. Even for the step
potential in this energy zone there will exist, during this transitory
time, current flow both {\em into} and then subsequently {\em out} of the
step. Since, the stationary solution has zero current flow within the step,
this feature is not always recognized. It is however obvious when one
admits that there will be a non zero transitory $\mbox{d}\rho/\mbox{d}t$
for any space interval within the step.

The most important barrier feature (both theoretically and experimentally)
of this energy zone is tunnelling. A part of the incoming wave packet will
continue its course beyond the barrier region. Its magnitude will be
modulated by the barrier. For "large" barriers (compared to the wave packet
size) an exponential reduction in amplitude $\propto \exp[-ql]$ occurs,
where $l$ is the barrier length and $q$ is
$\sqrt{m^{\2}-(E-V_{\0})^{^{\2}}}$. This not only reduces the amplitude but
modifies the transition spectrum. The smaller the transition amplitude is,
the smaller the modifications in the {\em reflected wave packet} from the
incident wave packet. However,in general, for both for the reflected and
transmitted wave packets we have maxima in configuration space, and can
apply the stationary phase method (SPM) to calculate reflection time delays
and transition times \cite{SPM}. With the Schr\"odinger equation the
conclusion that the transition time is {\em independent} of barrier width
$l$, when $l\to \infty$, is known as the Hartman effect \cite{HE}. Such a
result is hard to avoid and, if the same occurs for the Dirac equation
(subject matter of this paper and previously discussed by other authors
\cite{Kre1,Kre2,Pet}), we would have to face the unpalatable feature of
superluminal velocities within the barrier. We  warn that more than one
type of transition time has been defined in the literature
\cite{SPM,T1,T2,T3} and for details we refer to reference \cite{NH,Rec}. In
this paper, we intend to investigate this particular aspect of tunnelling
by means of the SPM  neglecting the possible ambiguities that this
approximation is known to have.

In the next section, we define all quantities and equations used.
Some of these have been given also elsewhere \cite{DifD,KleD} but
for completeness we re-present them. We also solve the stationary
plane wave problem for the step and barrier. In Section III, we
calculate the transition times by using the SPM and, based on
numerical calculations, we discuss the appearance of superluminal
velocities. Our conclusion are drawn in Section IV.

\section*{II FORMALISM AND SOLUTIONS FOR THE BARRIER}

We shall work with a one-dimensional (electrostatic) potential in
the Dirac equation. The chosen axis is the $z$-axis. However, we
shall  use the solutions and hence formalism of the full three
dimensional case \cite{Perskin}. Thus, the stationary Dirac
equation reads
\begin{equation}
\left[ \, E - V(z) \, \right] \, \gamma_{\0}\, \psi(z) +
i\,\gamma_{\3}\,\psi'(z) = m \, \psi(z)\,\,,
\end{equation}
where $\gamma_{\0}$ and $\gamma_{\3}$ are two of the Dirac matrices (see
below) and $\psi'(z)=\mbox{d}\psi(z)/\mbox{d}z$. The explicit time
dependence $\exp[-iEt]$ has been dropped from the above equation and hence
$\psi$ is only a function of the $z$ coordinate. Our chosen representation
for the gamma matrices is the Pauli-Dirac one, so that \begin{equation}
\gamma_{\0}=\left(
\begin{array}{rr} 1 & 0 \\ 0 & \,\,-1\end{array} \right)\,\,\, , \, \, \,
\, \, \gamma_{\3}=\left(
\begin{array}{cc} 0 & \sigma_{\3} \\ - \sigma_{\3} & 0 \end{array}
\right)\,\, .\end{equation}
The barrier potential is fixed at $V_{\0}$ in
the region $0<z<l$ and is zero elsewhere. The $z$ axis is divided into
three regions. Region I is the region of the incident and reflected waves
($z<0$). Region II is the barrier region. Region III is that containing the
transmitted wave ($z>l$).
For $V_{\0}>m$, it is convenient to divide the tunnel energy zone  into two
sub-zones both evanescent: (A) $V_{\0}<E<V_{\0}+m$ and (B)
$V_{\0}-m<E<V_{\0}$ ($V_{\0}>2m$) or $m<E<V_{\0}$ ($m<V_{\0}<2m$). For
$V_{\0}<m$, only the evanescent zone (A), $m<E<V_{0}+m$, exists. The A zone
corresponds to energy above potential ($E>V_{\0}$) and we will use the
"positive energy" $u^{(s)}$ solutions modified for a non zero potential
$V_{\0}$. The B zone corresponds to the below potential zone ($E<V_{\0}$).

One of the questions we pose in this work is if all below
potential solutions ("negative energy") represent physical
antiparticles, be they oscillatory (free) or evanescent. For the B
zone we shall use the $u^{(s\pl \2)}$ solutions modified to allow
for a constant non zero potential $V_{\0}$.

Spin flip is absent in all these problems (independent of the value of $E$)
so by choosing an incoming spin up state, \begin{equation} u^{(\1)}(p\,,E)=
[\,1\,\,,\,\,\,0\,\,,\,\,\,p/(E+m)\,\,,\,\,\,0\,]^t\,\,,\end{equation} we
find the following spinors in region II,
 \begin{equation}
\begin{array}{lclcl}
\mbox{\sc A-zone}: &\,\,\, & u^{(\1)}(\pm iq\,,E-V_{\0}) & = &
[\,1\,\,,\,\,\,0\,\,,\,\,\,\pm iq/(E-V_{\0}+m)\,\,,\,\,\,0\,]^t\,\,,\\
\mbox{\sc B-zone}: &  &  u^{(\3)}(\pm iq\,,|E-V_{\0}|) & =& [\,\mp
iq/(|E-V_{\0}|+m)\,\,,\,\,\,0\,\,,\,\,\,1\,\,,\,\,\,0\,]^t\,\,.\end{array}
\end{equation}
Only the spinor $u^{(\1)}(p\,,E)$ appears in region III.

\subsection*{A-ZONE: $\boldsymbol{V_{\0}<E<V_{\0}+m\,\,(V_{\0}>m)}$
OR $\boldsymbol{m<E<V_{\0}+m\,\,(V_{\0}<m)}$}

The solutions in the three regions are:
\begin{equation}
\begin{array}{lclcrcl}
 \mbox{\small \sc Region I:} &~~~~ & \hspace*{.65cm} z < 0 \, ,& ~~~ &
 \hspace*{.85cm}
u^{(\1)}(p\,,E)\,\exp[ipz] &+&
   R_{>}\,u^{(\1)}(-p\,,E)\,\exp[-ipz]\,\, , \\
\mbox{\small \sc Region II:} & &0 < z < \,l \, ,&  & A_{>}\,
u^{(\1)}(iq\,,E-V_{\0})\,\exp[-qz] &+&
   B_{>}\,u^{(\1)}(-iq\,,E-V_{\0})\,\exp[qz]\,\, , \\
\mbox{\small \sc Region III:} &   & \,l < z \, ,&  & T_{>}\,
u^{(\1)}(p\,,E)\,\exp[ipz] & ,&
\end{array}
\end{equation}
and we are using un-normalized solutions but such that
$|R_{>}|^{^{\2}}$ is the reflection probability. The symbolism
$R_{\lessgtr}$ ($T_{\lessgtr}$) will be used for the $A/B$ energy
zones because $E\lessgtr V_{\0}$ respectively. The solutions in
region II are the evanescent solutions $\propto \exp[\pm qz]$. We
shall in what follows refer to the case of a step potential with
only two regions (I and II), without treating this case separately
we merely note that it corresponds to the above solutions with
$B_{\lessgtr}=0$. It should also be obtainable from the barrier
solution when $l\to \infty$, although some care must be taken when
multiple peaks occur such as in the case of above potential
diffusion \cite{DifD}. The first of these barrier peaks reproduces
the single step peak in the $l\to \infty$ limit.

Solving the continuity equations, $\psi_{\I}(0)=\psi_{\II}(0)$ and
$\psi_{\II}(l)=\psi_{\III}(l)$, in matrix form yields
\begin{equation}
 \left(\, \begin{array}{rr} 1 & 1 \\
 1 & -1
 \end{array}\, \right)\,\left[\,\begin{array}{c}
1\\ R_{>}
\end{array}\,\right]=\left(\, \begin{array}{rr} 1 & 1 \\
 \alpha & \,-\alpha
 \end{array}\, \right)\,\left[\,\begin{array}{c}
A_{>} \\ B_{>}
\end{array}\,\right]\,\,\,\,\,\mbox{and}\,\,\,\,\,
\left(\, \begin{array}{rr} e^{-ql} & e^{ql} \\
 \alpha \,e^{-ql}& \,\,-\alpha \,e^{ql}
 \end{array}\, \right)\,\left[\,\begin{array}{c}
A_{>} \\ B_{>}
\end{array}\,\right] =\left[\,\begin{array}{r}
T_{>} \\ T_{>}
\end{array}\,\right]\,e^{ipl}\,\,,
\end{equation}
where
\[\alpha=\,i\,\frac{q}{p}\,
\frac{E+m}{E-V_{\0}+m}\,\,. \]
Since Dirac is a first order equation only
continuity of $\psi$ is imposed. Solving the above equations gives
\begin{eqnarray}
\left[\,\begin{array}{c} 1\\ R_{>}
\end{array}\,\right]&=&
\left(\, \begin{array}{rr} 1 & 1 \\
 1 & -1
 \end{array}\, \right)^{-\1}\,
\left(\, \begin{array}{rr} 1 & 1 \\
 \alpha & \,-\alpha
 \end{array}\, \right)\,
\left(\, \begin{array}{rr} e^{-ql} & e^{ql}\\
 \alpha \,e^{-ql}& \,\,-\alpha \,e^{ql}
 \end{array}\, \right)^{- \1}\,
\left[\,\begin{array}{r} T_{>} \\ T_{>}
\end{array}\,\right]\,e^{ipl} \nonumber \\
 & = &\mbox{$\frac{1}{2}$}\,\left[\, \begin{array}{rr} \cosh(ql) +
\alpha\, \sinh(ql) & \cosh(ql) + \frac{\sinh(ql)}{\alpha}\\
\cosh(ql) - \alpha \,\sinh(ql)  & \,\,\,\,\,- \cosh(ql) +
\,\frac{\sinh(ql)}{\alpha}
\end{array}\, \right]\, \left[\,\begin{array}{r} 1\\1
\end{array}\,\right]\,T_{>}\,e^{ipl}
\end{eqnarray}
Thus,
\begin{equation}
R_{>} = \frac{1-\alpha^{\2}}{2\alpha}\,\sinh(ql)\,\,T_{>}\,
\exp[\,i\,p\,l\,] \,\,\,\,\, \mbox{and} \,\,\,\,\,
 T_{>} =  \exp[\,- i\,p\,l\,]\,\mbox{\Large /}\left[\, \cosh (ql) +
\frac{1\,+\,\alpha^{\2}}{2\alpha} \,\sinh(ql)\,\right]\,\,.
\end{equation}
The non relativistic limit, $E-m=E_{\NR}\ll m$ and $V_{\0} \ll m$,
reproduces the Schr\"odinger results for the reflection and transmission
coefficients (see the appendix for a detailed derivation).

\subsection*{B-ZONE: $\boldsymbol{V_{\0}-m<E<V_{\0}\,\,(V_{\0}>2m)}$ OR
$\boldsymbol{m<E<V_{\0}\,\,(m<V_{\0}<2m)}$}

For this zone, in the potential region, we shall use the spinor
$u^{(\3)}$. Thus, we have
\begin{equation}
\begin{array}{lclcrcl}
 \mbox{\small \sc Region I:} &~~~~ & \hspace*{.65cm} z < 0 \, ,& ~~~ &
 \hspace*{.85cm}
u^{(\1)}(p\,,E)\,\exp[ipz] &+&
   R_{<}\,u^{(\1)}(-p\,,E)\,\exp[-ipz]\,\, , \\
\mbox{\small \sc Region II:} & &0 < z < \,l \, ,&  & A_{<}\,
u^{(\3)}(iq\,,|E-V_{\0}|)\,\exp[-qz] &+&
   B_{<}\,u^{(\3)}(-iq\,,|E-V_{\0}|)\,\exp[qz]\,\, , \\
\mbox{\small \sc Region III:} &   & \,l < z \, ,&  & T_{<}\,
u^{(\1)}(p\,,E)\,\exp[ipz] & ,&
\end{array}
\end{equation}
Continuity equations in matrix form yields
\begin{equation}
 \left(\, \begin{array}{rr} 1 & 1 \\
 1 & -1
 \end{array}\, \right)\,\left[\,\begin{array}{c}
1\\ R_{<}
\end{array}\,\right]=\mbox{$\frac{E+m}{p}$}\,\left(\, \begin{array}{rr} -\beta & \beta \\
 1 & 1
 \end{array}\, \right)\,\left[\,\begin{array}{c}
A_{<} \\ B_{<}
\end{array}\,\right]
\end{equation}
and
\begin{equation}
\mbox{$\frac{E+m}{p}$}\,\left(\, \begin{array}{rr} -\beta\,e^{-ql} & \,\,\beta\,e^{ql} \\
 e^{-ql}& \,\,e^{ql}
 \end{array}\, \right)\,\left[\,\begin{array}{c}
A_{<} \\ B_{<}
\end{array}\,\right] =\left[\,\begin{array}{r}
T_{<} \\ T_{<}
\end{array}\,\right]\,e^{ipl}\,\,,
\end{equation}
where
\[\beta=\,i\,\frac{qp}{(E+m)(|E-V_{\0}|+m)}\,\,. \]
Solving the above matrix equations, we find
\begin{eqnarray}
\left[\,\begin{array}{c} 1\\ R_{<}
\end{array}\,\right]&=&
\left(\, \begin{array}{rr} 1 & 1 \\
 1 & -1
 \end{array}\, \right)^{-\1}\,
\left(\, \begin{array}{rr}
 -\beta & \,\beta\\ 1 & 1
 \end{array}\, \right)\,
\left(\, \begin{array}{rr}
 -\beta \,e^{-ql}& \,\,\beta \,e^{ql}\\
 e^{-ql} & e^{ql}
 \end{array}\, \right)^{- \1}\,
\left[\,\begin{array}{r} T_{<} \\ T_{<}
\end{array}\,\right]\,e^{ipl} \nonumber \\
 & = &\mbox{$\frac{1}{2}$}\,\left[\, \begin{array}{rr} \cosh(ql) -\frac{\sinh(ql)}{\beta}
& \cosh(ql) - \beta\, \sinh(ql)\\
\cosh(ql) +\frac{\sinh(ql)}{\beta}  & \,\,\,\,\,- \cosh(ql) -
\,\beta\, \sinh(ql)
\end{array}\, \right]\, \left[\,\begin{array}{r} 1\\1
\end{array}\,\right]\,T_{>}\,e^{ipl}
\end{eqnarray}
Thus,
\begin{equation}
R_{<} =  \frac{1-\beta^{\2}}{2\beta}\,\sinh(ql)\,\,T_{<}\,
\exp[\,i\,p\,l\,] \,\,\,\,\, \mbox{and}\,\,\,\,\,
 T_{<} =  \exp[\,- i\,p\,l\,]\,\mbox{\Large /}\left[\, \cosh (ql)
 -\frac{1\,+\,\beta^{\2}}{2\beta} \,\sinh(ql)\,\right] \,\,.
\end{equation}
Although not obvious, simple algebraic calculations show that
$R_{<}$ and $T_{<}$ are functionally identical to $R_{>}$ and
$T_{>}$ respectively (although, of course, valid in disjoint
energy zones). Hence in the following, we will drop the suffixes
and use
\begin{eqnarray}
T &= & \exp\left\{ -ip\,l+i \arctan \left[
\frac{E^{^{\2}}-m^{\2}-EV_{\0}}{qp}\, \tanh(ql)
 \right] \right\} \mbox{\Large /} \sqrt{ \cosh^{\2}(ql) +
 \left[ \frac{E^{^{^{\2}}}-m^{^{\2}}-EV_{\0}}{qp}\,\sinh(ql)
 \right]^{^{\2}}} \,\,,\nonumber \\
 R & = & -\,i\,\frac{mV_{\0}}{qp}\,\sinh(ql)\,\,T\,
\exp[\,i\,p\,l\,]\,\, .
\end{eqnarray}

\section*{III. TIME ANALYSIS}

In this Section, we shall calculate the analytic expression of the
transition times by using the SPM and, by numerical calculation,
we present the appearance of superluminal velocities.

\subsection*{$\bullet$ SPM TRANSITION TIMES}

For this analysis the essential ingredient is the phase of the
transmitted amplitude $T$. The gaussian envelope function $g(p)$
will be assumed real and the wave packet function $\Psi(x,t)$,
defined in the standard way, is given by
\begin{equation}
\Psi(x,t)=\int \mbox{d}p\,g(p)\,\psi(z)\, \exp[\,-i\,E\,t)\,]\,\,.
\end{equation}
 A common choice for $g(p)$
(unnormalized) peaked at $p_{\0}$ is
\[ g(p)=\exp[-a^{\2}(p-p_{\0})^{\2}/4]\,\,,\]
so that for the incoming wave
$\psi_{\I,\inc}(z)=u^{(\1)}(p\,,E)\,\exp[ipz]$ the wave packet
width is just $a$ (large barriers are thus defined by $l/a\gg 1$).
Due to the real nature of $g(p)$, the phase in
$\Psi_{\I,\inc}(x,t)$ is simply
\begin{equation}
\phi_{\I,\inc} = p\,z-E\,t\,\,.
\end{equation}
The SPM then sets the maximum of the incident wave packet at time
$t$ at
\[ z =
\left[\frac{\mbox{d}E}{\mbox{d}p}\right]_{\0}t=\frac{p_{\0}}{E_{\0}}\,t\,\,,\]
where $E_{\0}$ is the energy corresponding to the peak momentum
value of $p_{\0}$.  Whence the incoming wave packet maximum
reaches (ignoring interference effects with the reflected wave)
$z=0$ at time $t=0$.

The SPM calculation of the transmission time uses the phase factor
of $T$ obtained in the previous Section, \begin{equation}
\phi_{\III,\tra} = \arctan \left[
\frac{E^{^{\2}}-m^{\2}-EV_{\0}}{qp}\, \tanh(ql)
 \right]+ p\,(z-l)-E\,t\,\,. \end{equation}
Deriving $\phi_{\III,\tra}$ with respect to $E$ and setting $z=l$,
we find the following functional
\begin{eqnarray}
\label{ttra} t_{\tra}(E,l)
         & = &\left\{ \frac{2E-V_{\0}}{qp}\,\left[1 +
 \left( \frac{E^{^{\2}}-m^{\2}-EV_{\0}}{qp}\right)^{^{\2}}\,\right]
 \tanh(ql) +
\frac{(E^{^{\2}}-m^{\2}-EV_{\0})(V_{\0}-E)\,l}{q^{\2}\,p\,\cosh^{\2}(ql)}
\right\}
 \mbox{\Large /} \nonumber \\
 & &  \mbox{\Large /} \left\{ 1 +
 \left[ \frac{E^{^{\2}}-m^{\2}-EV_{\0}}{qp}\, \tanh(ql)
 \right]^{^{\2}}\right\}\,\,.
         \end{eqnarray}
The exit time of a single transmitted wave packet is then given by
$t_{\tra}(\tilde{E}_{\0},l)$, where
$\tilde{E}_{\0}=\sqrt{\tilde{p}_{\0}^{\2}+m^{\2}}$ and $\tilde{p}_{\0}$ is
the peak momentum of the transmitted wave packet, i.e. (neglecting the
spinor momentum dependencies) the maximum of $g(p)\,|T|$ (see Fig.1 and
Fig.3). If the barrier is short then $\tilde{p}_{\0}\approx p_{\0}$ because
the gaussian dominates over the transmission amplitude. On the other hand,
if we take $l\to \infty$ the functional $t_{\tra}(E,l)$ greatly simplifies
giving
\begin{equation}
\tau_{\tra}(E):=\lim_{l\to
\infty}t_{\tra}(E,l)=\frac{2E-V_{\0}}{qp}\,\,,
\end{equation}
which is independent of $l$. At first glance,  this would mean
unlimited tunnelling velocities. Actually, this is often argued
without taking into account the difference between $p_{\0}$ and
$\tilde{p}_{\0}$. The exit time is given by $\tau(\tilde{E}_{\0})$
and not by $\tau(E_{\0})$. If for example we set $\tilde{E}_{\0}$
to its maximum allowed value (compatible with tunnelling)
$\tilde{E}_{\0}=V_{\0}+m$, we find
\[ \tau_{\tra}[V_{\0}+m] \to \infty\,\, ,\]
so that the tunnelling velocities are {\em not} in general
unlimited. However, since we shall find in the next Section
superluminal velocities for finite $l$, we shall not dwell upon
the asymptotic tunnelling velocity. We only note that if instead
of a gaussian wave packet (which technically overshoots the
tunnelling zone) we use a truncated gaussian, we can avoid
infinite SPM tunnelling times by truncating below $V_{\0}+m$. We
warn however that a truncation in the momentum spectrum of a wave
packet automatically introduces infinite wave packets in
configuration space and we have to take care in using the SPM
\cite{DifS,DifD}.


\subsection*{$\bullet$ SUPERLUMINAL VELOCITIES}

In the previous sub-Section, we derived an expression for the
transmission time,
\[ t_{\tra}(\tilde{E}_{\0},l)\,\, .\]
This expression requires the knowledge of $\tilde{E}_{\0}$, the peak value
of the transmitted momentum distribution. This implicitly assumes a single
maximum. So, the SPM is certainly valid for moderate values of $l$, where
the transmitted spectrum is almost gaussian as it is shown in Fig.1 and
Fig.3 ($l \lesssim 2a$). In such a context, we have numerically calculated
$\tilde{E}_{\0}(l)$ and whence obtained $t_{\tra}(\tilde{E}_{\0},l)$ by
Eq.(\ref{ttra}). In Fig.2-a and Fig.4-a, we have plotted
$t_{\tra}(\tilde{E}_{\0},l)/a$ against $l/a$. We observe that for $l\gg a$,
we have to  use instead of the maximum momentum value, the average value of
the spectrum, i.e. $\tilde{E}_{\0}\equiv \langle E\rangle$. This is because
for a momentum curve which ends upon a maximum, and it is thus very
asymmetric, model numerical calculations show that a more accurate result
for SPM times is obtained with the use of $\langle E\rangle$. The
surprising feature of the curves given in Fig.2-a and Fig.4-a is obtained
by taking the ratio of the coordinates,
\[ v_{\tra}(\tilde{E}_{\0},l)=\frac{l}{t_{\tra}(\tilde{E}_{\0},l)}\,\,,\]
the effective velocity of the tunnelling processes. This is plotted in
Fig.2-b and Fig.4-b as function of $l/a$. As it can be seen from Fig.2-a
and Fig.4-a, there is a plateau region where the transit time is
independent of the value of $l/a$. In this region the effective velocity
grows linearly (see Fig.2-b and Fig.4-b). However, the surprising feature
is the numerical value of the velocity in this region: it is already
greater than one both for the relativistic and for the non relativistic
case. We have no need to go to the infinite barrier width limit (Hartman
effect) to find superluminal velocities.



\section*{IV. CONCLUSIONS}

We have studied in this paper the tunnelling phenomena predicted by the
Dirac equation. One of the principle questions posed at the start was if a
Hartman effect exists also for the Dirac equation. The answer is positive
since the spinors play no significant role in the calculation of the
transmission times. There is a difficulty with the fact that the different
momentum dependence of the spinors lead to different transmission times for
the components. However, this does not modify the result of each exhibiting
a Hartman-like effect. It is the SPM which obliges us to work with wave
function {\em amplitudes} rather than with the transmission probability
function, in which spinor components have been summed over. However, if all
spinor components yield superluminal velocities then superluminal
velocities must be expected. The Hartman limit ($l \gg a$) has an added
complication because the transmission function dominates the incoming
convolution function (gaussian in all our calculations) and it is not yet
clear how, or even if, the SPM works in the absence of a clean maximum in
the momentum distribution. We have avoided entering into this equation
because there is no need to go to $l\gg a$ in order to exceed the velocity
of light.

A second question involved in our study was the identification of
the nature (charge) of the particles temporarily (for wave
packets) in the classically forbidden barrier region. This is a
relevant question when one recalls that in the Klein energy zone
($E<V_{\0}-m$) antiparticles are created and/or annihilated in the
barrier region. For an antiparticle the barrier becomes a well and
the {\em mathematical} below potential ($V_{\0}$) particles of
energy $E$ are re-interpreted as {\em physical} above potential
($-V_{\0}$) antiparticles of energy $-E$\cite{KleD,K3}. It is
tempting to consider all "particles" with $E<V_{\0}$ (below
potential) to be in fact physical antiparticles, even if
associated to evanescent terms (B tunnelling zone). This is the
reason, we divided the tunnelling energy zone into two the A
($V_{\0}<E<V_{\0}+m$) and the B ($V_{\0}-m<E<V_{\0}$) zones. We
now give an argument based upon our studies of the Klein zone that
says that this hypothesis is {\em not} true.

Let us consider in the following the simple step potential. The
Dirac equation conserves probabilities. How is this consistent
with pair creation in which more particles are reflected $R>1$
(Klein paradox) than are incident? Physically, total charge is
conserved but probability certainly is not. The answer is
suggested by the well know fact that the below potential particles
in the Klein zone have the "wrong" group velocity. This fact
incidentally concords with the Feynman-St\"uckelber conclusion
that such below potential particles must actually travel backwards
in time. Returning to our conundrum, we observe that in any formal
numerical calculation, which ignores the antiparticle
re-interpretation, the Klein paradox appears mathematically as the
particular solution to a problem in which at $t=-\infty$ we have
{\em two opposite moving wave packets}: the incident wave packet
at $z=-\infty$ and the below potential wave packet (of appropriate
size) at $z=+\infty$. When these two meet at time $t\sim 0$ at the
step discontinuity $z=0$ the continuity equations tell us that
they unite and form a single wave packet, the reflected wave
packet \cite{AJP}. In this way probability is indeed seen to be
conserved. It is the re-interpretation in physical
particle/antiparticle terms which alters our viewpoint. However,
the Dirac equation (with a real potential) can be viewed in this
mathematical picture with {\em only particles}, albeit with
energies both above and below potential. Indeed this is the way
everyone treats the stationary plane wave problem, including
ourselves \cite{KleD}.

Let us now apply the same mathematical viewpoint to the step in
the energy zones A and B. In this case there cannot be any
effective particle flow from $z=\infty$ since the stationary
solution under the barrier is, in both energy zones, a pure
exponential decreasing space function. At time $t=-\infty$ only
the incoming wave packet exists in this case. Eventually, for
$t=+\infty$ only the reflected wave packet exists. For times
within the transmission period during which complete reflection
occurs, we cannot have a reflection coefficient $R>1$, even if
only for an instant, without violating probability conservation.
This conclusion is independent of the choice of the A or B zones.
It is based upon the impossibility of having a modification of the
initial conditions so as to reproduce a transitory Klein-like
paradox. Consequently, even in tunnelling, the probability density
under the potential must represent the same particles as those of
the incoming wave.


It would be desirable to conclude the debate on superluminal velocities in
tunnelling phenomena but, at the moment, the results are far from being
conclusive. For a potential of the order of the mass, the Dirac equation
cannot be viewed as a one-particle equation, and particle creation is
expected to play an important role. There is still much to be studied in
potential problems within field theory and the final answer can only be
reached by analyzing tunnelling phenomena in a second quantized
theory\cite{Perskin,Gross}. However, this topic exceeds the scope of this
paper and it will be appropriately discussed in a forthcoming article. In
such a spirit, this paper has to be seen as an initial work to stimulate
further investigations. \\

\noindent{\bf ACKNOWLEDGMENTS}. One of the authors (SDL) wishes to thank
the Department of Physics of the Lecce University, where the paper was
prepared, for invitation and hospitality. He also thanks the FAEP (Brazil)
and the INFN (Italy) for financial support. The authors gratefully
acknowledge the helpful suggestions and comments of an anonymous referee
which allowed to improve the presentation of this paper. In particular,
they express their gratitude to the referee for drawing the attention to
the analysis of potentials much smaller than the mass of the particle. This
stimulated, in the revised version, a careful and detailed analysis of the
non relativistic limit.

\newpage

\section*{APPENDIX}

Let us discuss in detail the non relativistic (NR) limit, $E-m=E_{\NR}\ll
m$ and $V_{\0} \ll m$. We recall that for $V_{\0}<m$ only the evanescent
zone (A), $0<E_{\NR}<V_{\0}$, exists. For the convenience of the reader, we
rewrite the Dirac reflection and transmission coefficients given in the
text,
\begin{eqnarray}
T & = &  \exp[\,- i\,p\,l\,]\,\mbox{\Large /}\left[\, \cosh (ql) +
\frac{1\,+\,\alpha^{\2}}{2\alpha} \,\sinh(ql)\,\right]\,\,, \nonumber \\
R & =& \frac{1-\alpha^{\2}}{2\alpha}\,\sinh(ql)\,\, \exp[\,i\,p\,l\,] \,\,,
\end{eqnarray}
where
\begin{eqnarray}
p & \,=\, &\sqrt{E^{^{\2}}-m^{\2}} \,\, , \nonumber \\
q  & = &  \sqrt{m^{\2}-(E-V_{\0})^{^{\2}}}\,\, ,\\
\alpha & = & i\,\frac{q}{p}\, \frac{E+m}{E-V_{\0}+m}\,\, . \nonumber
\end{eqnarray}
Taking the NR limit, we obtain
\begin{eqnarray}
p & \,\to\, &p_{\NR}=\sqrt{2m\,E_{\NR}} \,\, , \nonumber \\
q  & \to &  q_{\NR}=\sqrt{2m\,(V_{\0}-E_{\NR})}\,\, ,\\
\alpha & \to & i\,q_{\NR}\,/\,p_{\NR}\,\, . \nonumber
\end{eqnarray}
Consequently,
\begin{eqnarray}
 T_{\NR} & = &  \exp[\,- i\,p_{\NR}\,l\,]\,\mbox{\Large /}\left[\,
 \cosh(q_{\NR} \,l) - i\,
 \frac{2E_{\NR}- V_{\0}}{2\,\sqrt{E_{\NR}(V_{\0}-E_{\NR})}}
  \,\sinh(q_{\NR} \,l)\,\right]\,\,,
  \nonumber \\
  R_{\NR} & =& -\,i\,\frac{V_{\0}}{2\,\sqrt{E_{\NR}(V_{\0}-E_{\NR})}}
\,\sinh(q_{\NR} \,l)\,\,T_{\NR} \exp[\,i\,p_{\NR}\,l\,] \,\,.
\end{eqnarray}
The square modulus of these coefficients,
\begin{eqnarray}
\left|T_{\NR}\right|^{\2} & = &
\frac{4\,E_{\NR}(V_{\0}-E_{\NR})}{4\,E_{\NR}(V_{\0}-E_{\NR})+
V_{\0}^{^{\2}} \sinh^{\2}\left[\sqrt{2m\,(V_{\0}-E_{\NR})} \,l\right]}
\,\,, \nonumber \\
\left|R_{\NR}\right|^{\2} & =&
\frac{V_{\0}^{^{\2}}}{4\,E_{\NR}(V_{\0}-E_{\NR})}
\,\sinh^{\2}\left[\sqrt{2m\,(V_{\0}-E_{\NR})}
\,l\right]\,\,\left|T_{\NR}\right|^{\2}\,\,,
\end{eqnarray}
is often encountered in standard quantum mechanics textbooks (see for
example \cite{Cohen} pag.\,73).

\newpage

\begin{figure}[hbp]
\vspace*{-1.5cm}\hspace*{-2.5cm}
\includegraphics[width=19cm, height=22cm, angle=0]{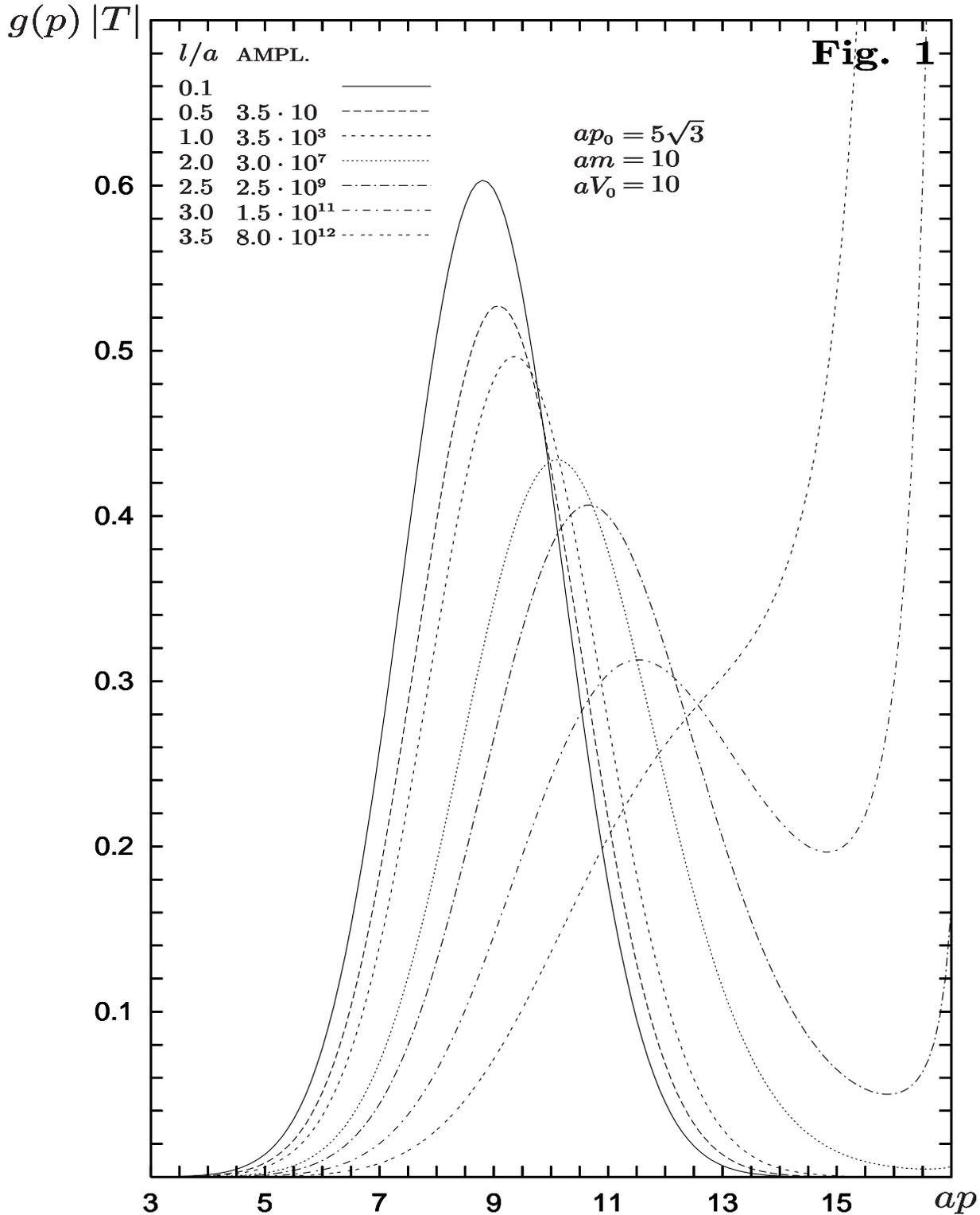}
\vspace*{-.5cm} \caption{The transmitted momentum distribution is plotted
as function of $ap$\, for different values of $l/a$, where $a$ is the width
of the incident wave packet at $t=0$. The potential is equal to the mass of
the particle, $aV_{\0}=am=10$, and the peak of the incident momentum
distribution is chosen to coincide with the center of the allowed zone
(compatible with tunnelling)  for the momentum, $ap_{\0}=a
\sqrt{V_{\0}(V_{\0}+2m)}/2=5\sqrt{3}$. For moderate values of $l/a$ the
transmitted momentum distribution is almost gaussian. The amplifications
show the attenuation (due to the evanescent waves) of the transmission
probability for increasing values of $l/a$.}
\end{figure}

\newpage

\begin{figure}[hbp]
\vspace*{-1.5cm}\hspace*{-2.5cm}
\includegraphics[width=19cm, height=22cm, angle=0]{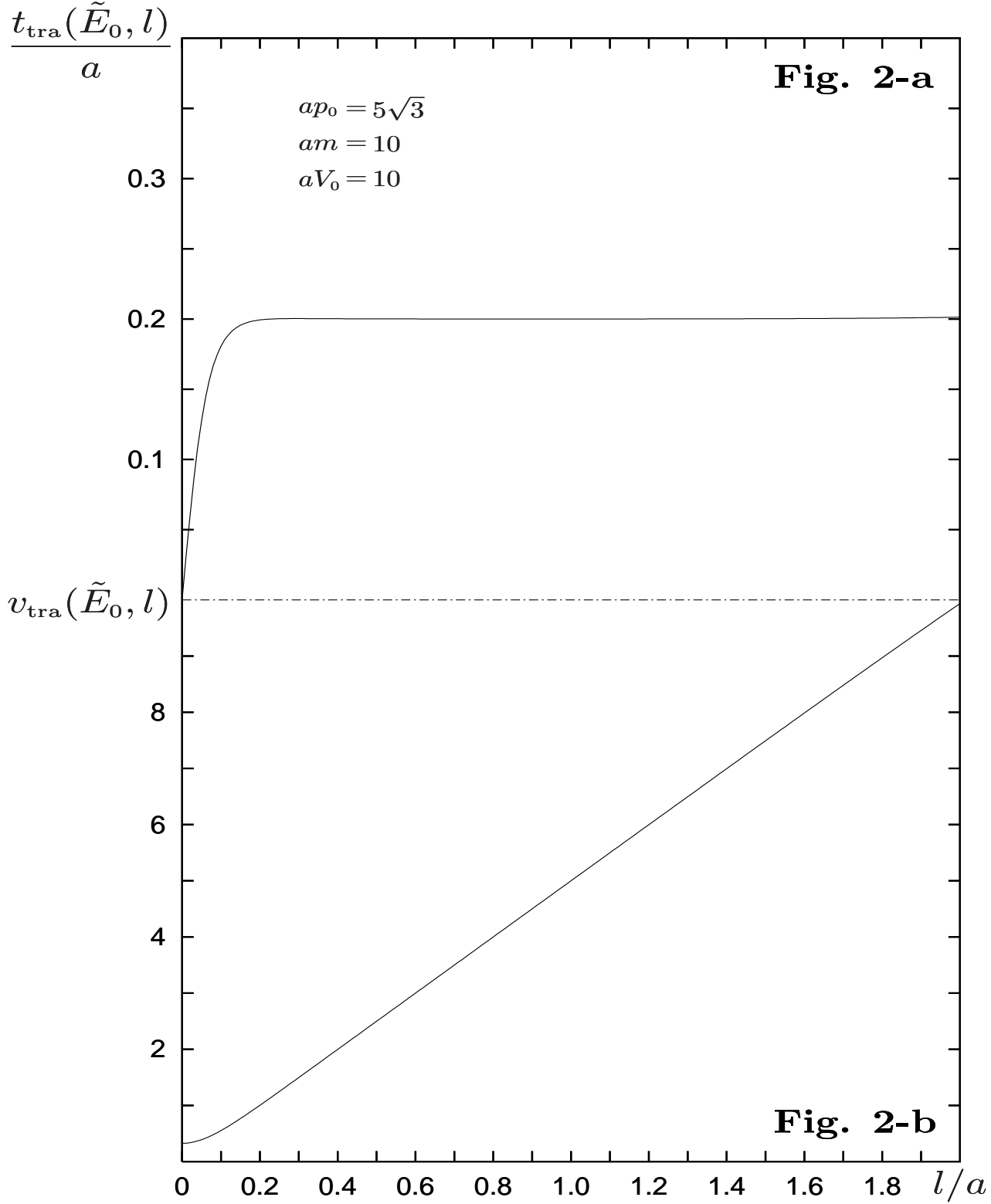}
\vspace*{-.5cm} \caption{This figure contains two curves. The plot in
Fig.2-a represents the variation of the (adimensional) transmission time,
$t_{\tra}(\tilde{E}_{\0},l)/a$ as function of $l/a$. The plot in Fig.2-b,
represents the ration between the barrier width, $l$, and the transmission
time $t_{\tra}(\tilde{E}_{\0},l)$, i.e. the effective velocity of the
tunnelling process. The potential is equal to the mass of the particle,
$aV_{\0}=am=10$, and the peak of the incident momentum distribution is
$ap_{\0}= 5\sqrt{3}$. The maximum value of $l/a$ has been chosen to be
$2.0$ in order to have an almost gaussian transmitted wave packet. This
guarantees the validity of the SPM. The surprising feature of our numerical
analysis is that the tunnelling velocity is already greater than one for a
moderate barrier width.}
\end{figure}

\newpage

\begin{figure}[hbp]
\vspace*{-1.5cm}\hspace*{-2.5cm}
\includegraphics[width=19cm, height=22cm, angle=0]{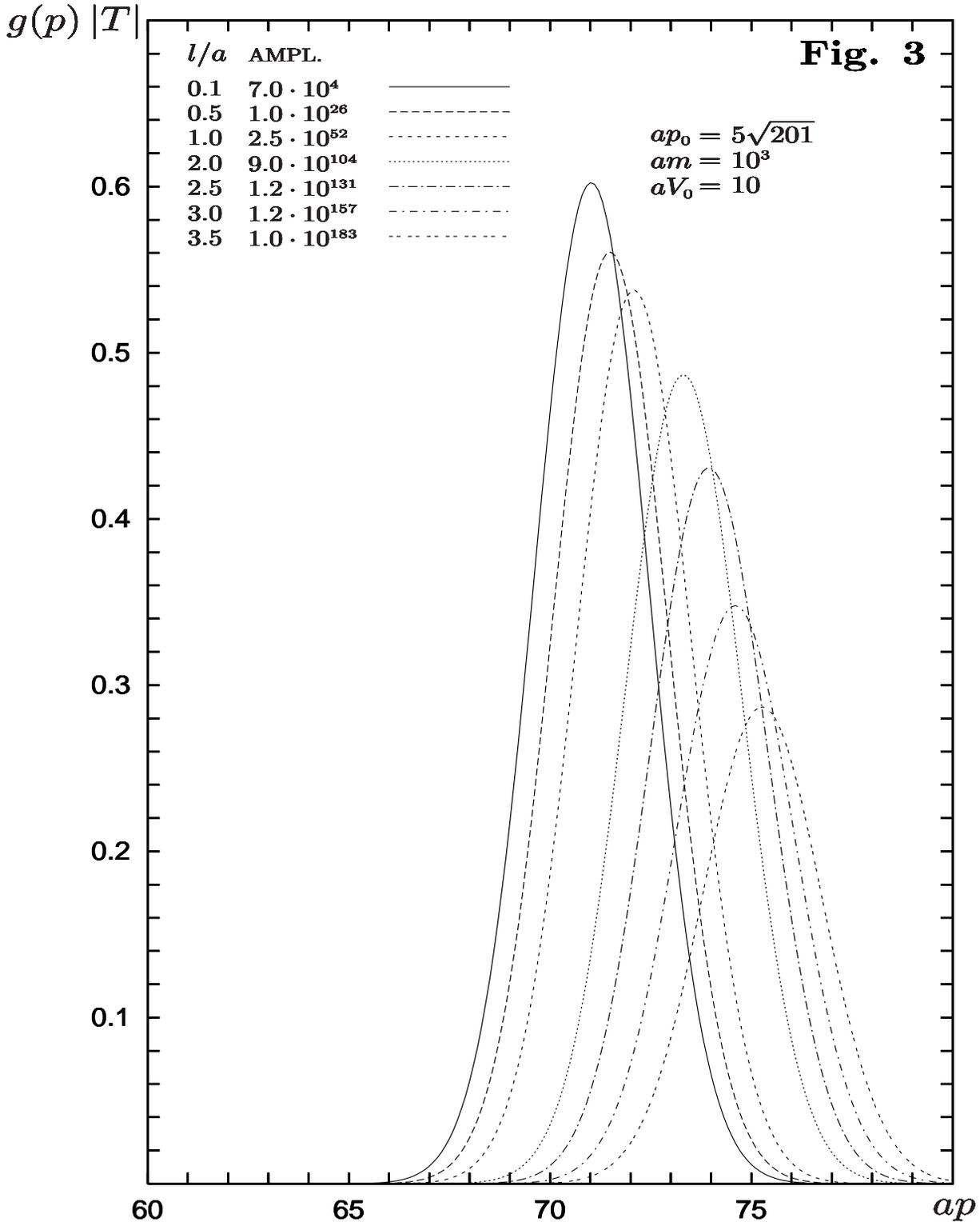}
\vspace*{-.5cm} \caption{The transmitted momentum distribution is plotted
as function of $ap$\, for different values of $l/a$. The potential
$aV_{\0}=10$ is now much smaller than the mass of the particle,
$am=10^{\3}$, and the peak of the incident momentum distribution is yet
chosen to coincide with the center of the allowed zone for the momentum,
$ap_{\0}=
 am \sqrt{V_{\0}(V_{\0}+2m)}/2=5\sqrt{201}$. This case  represents the non
 relativistic limit, $V_{\0}\ll m$ and $E-m=E_{\NR}\ll m$. The amplifications show
 a considerable attenuation of the transmission probability with respect to the
 relativistic case.}
\end{figure}

\newpage

\begin{figure}[hbp]
\vspace*{-1.5cm}\hspace*{-2.5cm}
\includegraphics[width=19cm, height=22cm, angle=0]{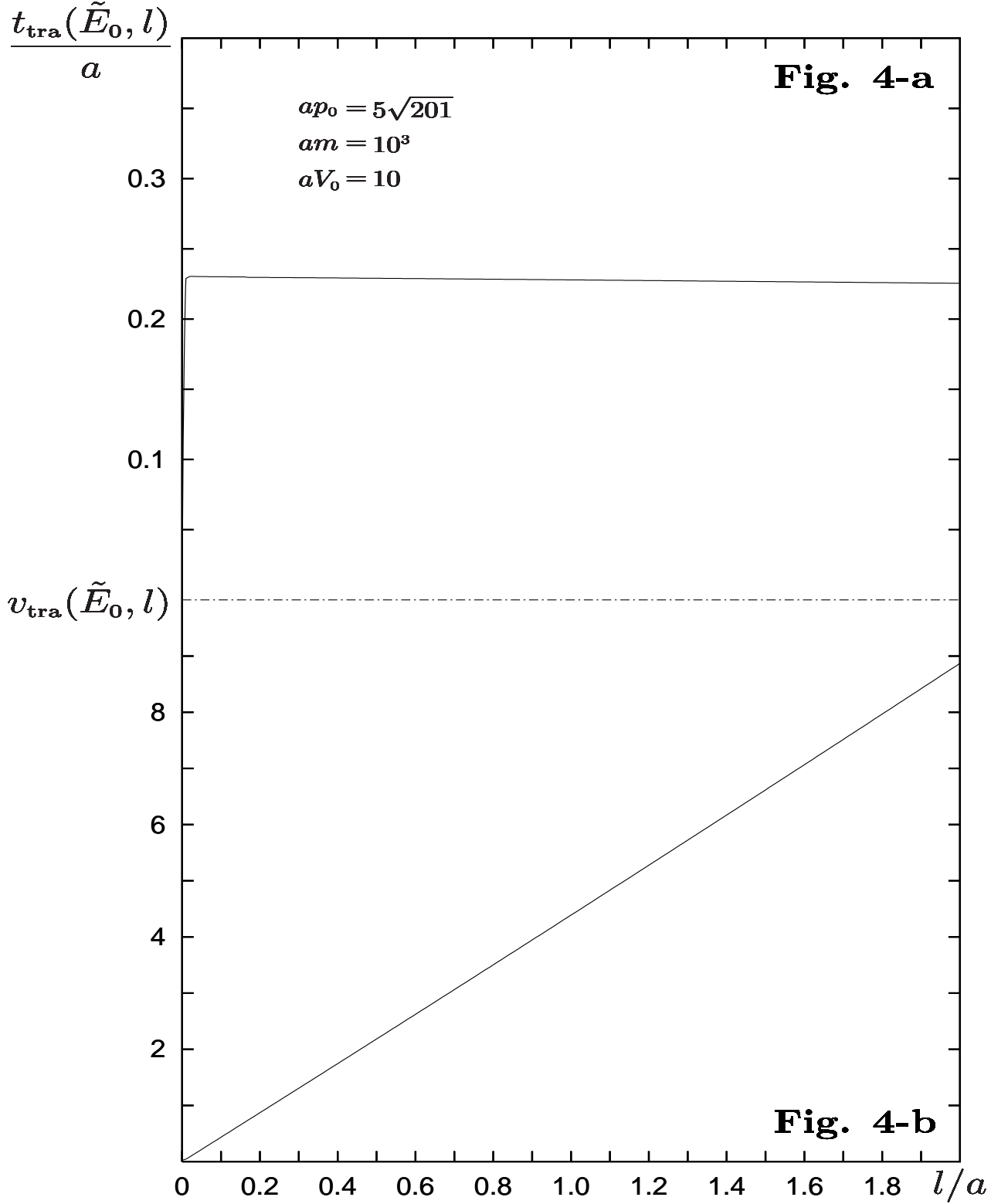}
\vspace*{-.5cm} \caption{The plot in Fig.4-a represents the variation of
the (adimensional) transmission time, $t_{\tra}(\tilde{E}_{\0},l)/a$ as
function of $l/a$. The plot in Fig.4-b, represents the ration between the
barrier width, $l$, and the transmission time $t_{\tra}(\tilde{E}_{\0},l)$,
i.e. the effective velocity of the tunnelling process. The potential
$aV_{\0}=10$ is much smaller than the mass of the particle, $am=10^{\3}$,
and the peak of the incident momentum distribution is $ap_{\0}=
5\sqrt{201}$. The maximum value of $l/a$ has been chosen to be $2.0$ in
order to have an almost gaussian transmitted wave packet. This guarantees
the validity of the SPM. The surprising feature of our numerical analysis,
i.e. the tunnelling velocity greater than one for a moderate barrier width,
is confirmed in the non relativistic limit.}
\end{figure}

\end{document}